\documentclass[aip,rsi,reprint]{revtex4-1}

\usepackage{graphicx}% Include figure files
\usepackage{dcolumn}% Align table columns on decimal point
\usepackage{bm}% bold math
\usepackage{url}

\begin{document}

%\preprint{RSI/CBT-DMZ120606}

%\preprint{}

\title{Metallic Coulomb Blockade Thermometry down to 10\,mK and below} %Title of paper



\author{L.~Casparis}
%\thanks{}
\affiliation{Department of Physics, University of Basel, CH-4056
Basel, Switzerland}

\author{M.~Meschke}
%\thanks{}
%\affiliation{Low Temperature Laboratory, Aalto University, School of Science, P.O. Box 13500, 00076 Aalto, Finland}
\affiliation{Low Temperature Laboratory, Aalto University, 00076 Aalto, Finland}

\author{D.~Maradan}
%\thanks{}
\affiliation{Department of Physics, University of Basel, CH-4056 Basel, Switzerland}

\author{A.~C.~Clark}
%\thanks{}
\altaffiliation{Stanford University, Department of Geophysics, Stanford, USA} \affiliation{Department of Physics,
University of Basel, CH-4056 Basel, Switzerland}

\author{C.~P.~Scheller}
%\thanks{}
\affiliation{Department of Physics, University of Basel, CH-4056 Basel, Switzerland}

\author{K.~K.~Schwarzw\"alder}
%\thanks{}
\affiliation{Department of Physics, University of Basel, CH-4056 Basel, Switzerland}

\author{J.~P.~Pekola}
%\thanks{}
\affiliation{Low Temperature Laboratory, Aalto University, 00076 Aalto, Finland}

\author{D.~M.~Zumb\"uhl}
\email[]{Dominik.Zumbuhl@unibas.ch}
%\thanks{}
\affiliation{Department of Physics, University of Basel, CH-4056 Basel, Switzerland}


\date{\today}

\begin{abstract}
We present an improved nuclear refrigerator reaching 0.3\,mK, aimed at microkelvin nanoelectronic experiments, and use
it to investigate metallic Coulomb blockade thermometers (CBTs) with various resistances $R$. The high-$R$ devices cool
to slightly lower $T$, consistent with better isolation from the noise environment, and exhibit electron-phonon cooling
$\propto T^5$ and a residual heat-leak of 40\,aW. In contrast, the low-$R$ CBTs display cooling with a clearly weaker
$T$-dependence, deviating from the electron-phonon mechanism. The CBTs agree excellently with the refrigerator
temperature above 20\,mK and reach a minimum-$T$ of $7.5\pm0.2$\,mK.
\end{abstract}


\maketitle %\maketitle must follow title, authors, abstract and \pacs


Advancing to ever lower temperatures can open the door for the discovery of new physics: for example, submillikelvin
temperatures in quantum transport experiments could lead to novel nuclear-spin physics\cite{Simon2007, Simon2008} in
nanoscale semiconductor devices \cite{HansonRMP} or could facilitate the study of non-Abelian anyons, Majorana Fermions
and topological quantum computation in fractional quantum Hall samples\cite{NayakRMP, SternRev}. However, cooling of
nanoscale devices below $T\sim1\,$mK is a formidable challenge due to poor thermal contact as well as microwave and
other heating, often resulting in device and/or electron temperatures raised well above the refrigerator temperature.
Therefore, significant progress beyond the status quo in both cooling techniques and thermometry is necessary.

One approach to overcome these difficulties uses Ag sinters\cite{lounasmaa,Pobell,Pickett} to thermalize the sample
wires\cite{SamkharadzeRSI}, pioneered by the Florida group\cite{Pan1999,Huang2007}. Another approach -- pursued by our
Basel group\cite{Clark2010} -- is to use nuclear cooling\cite{lounasmaa,Pobell,Pickett} on the sample wires, with the
potential to advance well into the microkelvin range. Thermometry in this regime\cite{lounasmaa,Pobell,Pickett}
typically faces similar challenges as cooling nanostructures and is ideally integrated on-sample. Among numerous
sensors\cite{Spietz2006}, Coulomb blockade thermometers\cite{Pekola1994} (CBTs) are simple to use and self-calibrating
yet offer high accuracy, demonstrated down to $\sim20\,$mK\cite{Meschke2010}. Here, we present an improved nuclear
refrigerator for cooling nanoelectronic samples and use it to investigate CBTs and their mechanisms of cooling.

\begin{figure}
\includegraphics[width=8.2cm]{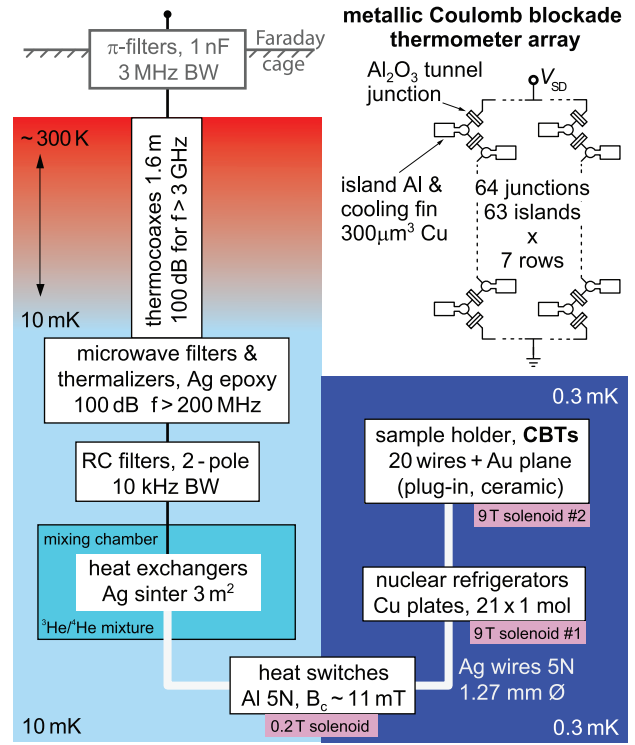}
\vspace{-2mm}\caption{\label{fig:Scheme} Layout of novel nanosample microkelvin refrigerator and CBT array. Radiation
shields (not drawn) are attached to the still and cold pate ($\sim50\,$mK). The RC filters are
$820\,\mathrm{\Omega}\,/\,22\,$nF and $1.2\, \mathrm{k\Omega}\,/\,4.7\,$nF. The 21 NR plates are $0.25\,\times
3.2\,\times 9.0\,\mathrm{cm^3}$ each, amounting to $64\,$g Cu per plate. }\vspace{-7mm}
\end{figure}

We employ a novel scheme for cooling electronic nano\-structures into the microkelvin regime by thermalizing each
sample wire directly to its own, separate nuclear refrigerator (NR) \cite{Clark2010}. In this scheme, the sample cools
efficiently through the highly conducting wires via electronic heat conduction, bypassing the phonon degree of freedom
since it becomes inefficient for cooling at low $T$. A prototype of this refrigerator presented in
Ref.\,\onlinecite{Clark2010} has been significantly improved in a 2$\mathrm{^{nd}}$ generation system, briefly outlined
below and in Fig.\,\ref{fig:Scheme}. A network of 21 parallel NRs is mounted on a rigid tripod intended to minimize
vibrational heating. Two separate $9\,$T magnets allow independent control of the NR and sample magnetic field.

Several stages of thermalization and filtering are provided on each sample wire (see Fig.\,\ref{fig:Scheme}). After
$\mathrm{\pi}$-filters and thermocoax\cite{Zorin}, each lead passes through a Ag-epoxy microwave filter
\cite{Scheller2011}, followed by an RC filter. Each wire then feeds into a Ag-sinter in the mixing chamber (MC),
emerging as a massive high-conductivity Ag wire. After Al heat-switches with fused joints, each lead traverses a
separate Cu NR via spot-welded contacts, terminating in an easily-exchangeable chip-holder plugged into Au-plated pins
which are spot welded to the Ag wires. Therefore, excellent thermal contact ($<50\,\mathrm{m\Omega}$) is provided
between the bonding pads and the parallel network of 21 Cu pieces -- the microkelvin bath and heart of the nuclear
refrigerator -- while maintaining electrical isolation of all wires from each other and from ground, as required for
nanoelectronic measurements.

The performance of the NRs is evaluated in a series of demagnetization runs. The temperature $T_{Cu}$ of the Cu pieces
is obtained using a standard technique\cite{Pobell, Pickett, Clark2010}: after demagnetization, we apply power on
heaters mounted on some of the NRs and evaluate the warm-up time-dependence $T_{Cu}(t)$ measured with LCMN thermometers
above $2\,$mK. This allows us to determine both the temperature $T_{Cu}$ of the Cu-NRs after demagnetization as well as
a small field-offset. For each demagnetization run, the NRs are precooled to $T_i\sim 12\,$mK in a $B_i=9\,$T magnetic
field and then demagnetized to temperatures as low as $T_f\sim0.3$ mK after the field has been slowly ramped down to
$B_f\sim 0.135\,$T, giving efficiencies $T_i/T_f\div B_i/B_f\gtrsim60\%$. Reruns showed excellent repeatability,
allowing us to chart $T_{Cu}$ for various $B_f$. To determine $T_{Cu}$ during the CBT experiments, we use the LCMN
thermometers above $2\,$mK, warm-up curves at the lowest $B_f$ and in-between, the pre-charted $T_{Cu}$ values.

\begin{figure}\vspace{-2mm}\hspace*{-2mm}
%\begin{figure}\hspace*{-5mm}\vspace{-2mm}
%\includegraphics[width=9.2cm]{f2.eps}
\includegraphics[width=9.4cm]{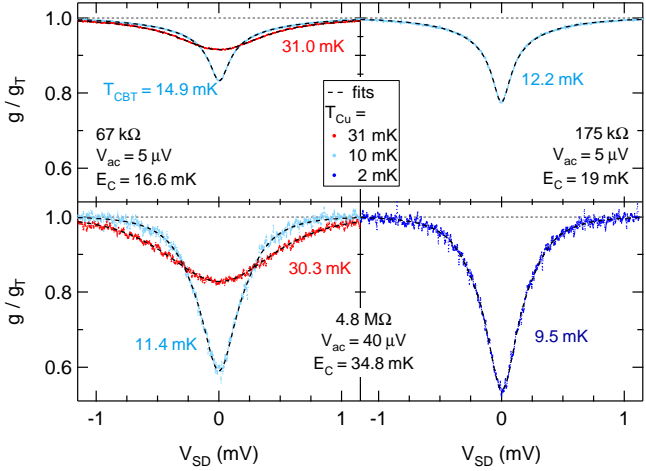}
\vspace{-7mm}\caption{\label{fig:gCBT} CBT normalized differential conductance $g/g_T$ versus source-drain dc bias
$V_{SD}$ for various NR temperatures $T_{Cu}$ as color-coded, with resulting $T_{CBT}$ ($\delta g$ method, see text)
given adjacent to each trace. Data from a $67\,\mathrm{k\Omega}$, a $175\,\mathrm{k\Omega}$ and a
$4.8\,\mathrm{M\Omega}$ CBT is shown. Dashed curves are fits to a model (see text). Note lower noise in low-$R$ sensors
due to larger resulting currents. }\vspace{-5mm}
\end{figure}

The network with 21 NRs allows measurements of several CBTs (2-wire each). The CBT devices are Au-wire bonded and glued
to the Au backplane of the chip carrier which is also cooled with a NR. Each CBT consists of 7 parallel rows of 64
Al/Al$_2$O$_3$ tunnel-junctions in series with an area of $2\,\mathrm{\mu m^2}$ fabricated using e-beam lithography and
shadow evaporation. The process used allows oxidation at elevated temperatures, giving junction resistances up to
$1\,\mathrm{M\Omega\,\mu m^2}$. Each island extends into a large cooling fin made from Cu, since Cu gives excellent
electron-phonon coupling. A small $B\sim 150\,$mT is applied perpendicular to the sensor wafer to suppress the
superconductivity of the Al. The differential conductance through a CBT sensor was measured with a standard lock-in
technique adding a small ac excitation $V_{ac}$ to a dc bias $V_{SD}$. Note that only $1/64$ of the applied voltage
drops across each junction and the sensor resistance $R$ is $64/7$ times the junction resistance $R_j$, assuming
identical junctions.

We investigated CBTs with various $R$, see Fig.\,\ref{fig:gCBT}. Due to Coulomb blockade effects, the conductance
around zero bias $V_{SD}=0$ is suppressed below the large-bias conductance $g_T$. Both width and depth $\delta
g=1-g(V_{SD}=0)/g_T$ of the conductance dip are related to the CBT electron temperature $T_{CBT}$. To extract
$T_{CBT}$, we perform fits (dashed curves) using a numerical model from Ref.\,\onlinecite{Meschke2004}. We find
excellent agreement between model and data (see Fig.\,\ref{fig:gCBT}). Independently, $T_{CBT}$ can be
obtained\cite{Meschke2004} from the conductance dip $\delta g=u/6-u^2/60+u^3/630$ with $u=E_C/(k_B T_{CBT})$ and
charging energy $E_C$. We first extract $E_C$ at high-$T$ assuming $T_{Cu}=T_{CBT}$ and then use this $E_C$ to extract
$T_{CBT}$ from $\delta g$ everywhere. While both methods produce very similar $T_{CBT}$ (deviating slightly only at the
lowest $T$), the $\delta g$ approach makes no a-priori assumptions about the cooling mechanism, allowing us an unbiased
investigation, though now requiring high-$T$ calibration against another thermometer (CMN). All $T_{CBT}$ values given
here are from the $\delta g$ method.

\begin{figure}[tr]\hspace*{-0mm}\vspace{-2mm}
\includegraphics[width=9.3cm]{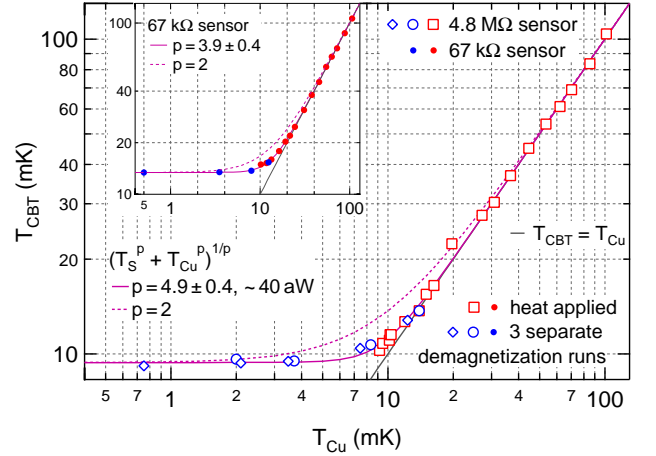}
\vspace{-8mm}\caption{\label{fig:f3} CBT electron temperature $T_{CBT}$ versus NR temperature $T_{Cu}$ for
$4.8\,\mathrm{M\Omega}$ (open markers) and $67\,\mathrm{k\Omega}$ sensors (filled markers, same axes on inset as main
figure). Below $10\,$mK, the data is obtained in 3 demagnetization sweeps (blue markers) with $B=9\,$T, $5\,$T, $2\,$T,
$1\,$T and $0.4\,$T in a typical run, ramped at $1\,$T/h above $1\,$T and $0.5\,$T/h below. Error bars are about the
size of the markers. Purple curves are $T_{CBT}$ saturation curves (see text). }\vspace{-4mm}
\end{figure}

The thermalization properties of $T_{CBT}$ of the lowest and highest $R$ CBTs are displayed in Fig.\,\ref{fig:f3} for a
wide range of $T_{Cu}$ from $0.5\,$mK to $100\,$mK. As seen, excellent agreement is found between $T_{CBT}$ and
$T_{Cu}$ at high temperatures, as expected. Further, $T_{CBT}$ is seen to lie well above $T_{Cu}$ at the lower
temperatures (see Fig.\,\ref{fig:gCBT} and \ref{fig:f3}), decoupling fully from $T_{Cu}$ well below $10\,$mK. We note
that $V_{ac}$ was experimentally chosen to avoid self heating. Also, the $4.8\,\mathrm{M\Omega}$ sensor reaches lower
temperatures than the other, lower impedance CBTs, consistent with better isolation from the environment, since the
power dissipated is proportional to $V_{env}^2/R_j$, with environmental noise voltage $V_{env}$.

To model the CBT thermalization\cite{Meschke2004}, we write down the heat flow $\dot{Q_i}$ onto a single island $i$
with electron temperature $T_i$:
\begin{equation}
\dot{Q_i}=\frac{V_j^2}{R_j}+\sum_{\pm}{\frac {\pi^2 k_B^2}{6 e^2 R_j}(T_{i\pm
1}^2-T_i^2)}-\Sigma\Omega(T_i^5-T_{p}^5)+\dot{Q_0} \label{heat}
\end{equation}
\noindent where $\dot{Q_0}$ is a parasitic heat leak and $V_j$ is the voltage drop across the junction, appearing here
in the Joule heating term. $\Sigma$ is the Cu electron-phonon (EP) coupling constant, $\Omega=300\,\mathrm{\mu m^3}$
the island volume and $T_{p}$ the phonon bath temperature assumed to be equal to $T_{Cu}$. This is well justified by
the high thermal conductance between the NRs and bonding pads. Note that at $T\ll 1\,$K, the sample-to-Au-backplane
interface resistance (Kapitza) is small compared to the EP coupling resistance \cite{Meschke2004}. Within this model,
two cooling mechanisms are included: Wiedemann-Franz (WF, $T^2$ term) and EP cooling. Note the strong $T^5$ dependence
of the EP term, ultimately rendering WF cooling dominant at sufficiently low $T$. Assuming one mechanism and
simplifying to only one island gives a saturation curve $T_{CBT}=(T_S^p+T_{Cu}^p)^{1/p}$, with a CBT saturation
temperature $T_S$ and an exponent $p$, corresponding to $p=2$ for WF-electron cooling and $p=5$ for EP cooling.

We study the mechanism of thermalization by fitting the saturation curve first to the $4.8\,\mathrm{M\Omega}$ data. We
find very good agreement, giving $p=4.9\pm0.4$ (see Fig.\,\ref{fig:f3}), indicating that EP coupling presents the
dominant cooling mechanism, limiting $T_{CBT}$ to $9.2\,$mK even though $T_{Cu}=0.75\,$mK. Using
$\dot{Q_0}=\Sigma\Omega T_{CBT}^5$, a small parasitic heat leak $\dot{Q_0}= 40\,$aW results for each island, with
$\Sigma=2\times10^9\,\mathrm{Wm^{-3}K^{-5}}$ from Ref.\,\onlinecite{Meschke2004}. We speculate that $\dot{Q_0}$ could
be caused by electrical noise heating such as microwave radiation, intrinsic residual heat release from materials used
or other heat sources. Considering the high-$R$ junctions and correspondingly weak WF cooling, it is not surprising
that EP coupling is dominant here.

When analogously examining the low-$R$ sensors, on the other hand, we find $p=3.9\pm0.4$ and $T_S=13.4\,$mK for the
$67\,\mathrm{k\Omega}$ sensor (see inset Fig.\,\ref{fig:f3}), and even $p=2.7\pm0.2$ and $T_S=6.9\pm0.1\,$mK for a
$134\,\mathrm{k\Omega}$ sensor (not shown) mounted on a conventional dilution refrigerator (base-$T\sim5\, $mK) with
improved filtering and chip-holder. Note that $T_S$ is the extrapolated $T=0$ saturation-$T$. The lowest $T$ measured
here was $7.5\pm0.2\,$mK. These power-laws far below $p=5$ indicate that EP cooling is no longer dominant but, rather,
a more efficient mechanism $p<5$ takes over at the lowest-$T$ in the low-$R$ sensors.

In summary, we have demonstrated operation of the NRs down to $0.3\,$mK while the CBTs cool as low as $7.5\,$mK. Though
the high-$R$ sensor is obviously cooled by EP coupling, the low-$R$ sensors, interestingly, appear to be entering a
different cooling regime. However, the low-$R$ sensors have slightly higher $T_{CBT}$ given the same environment,
consistent with stronger coupling to the environment. The lowest CBT temperatures are limited by the parasitic heat
leak, which is drained by the cooling channels available.

To further improve the sensor performance, the cooling-fin volume can be increased or the heat leak can be reduced,
potentially using improvements in microwave shielding and filtering, e.g. using on-chip capacitors, me\-tal planes or
alternative array designs. Such efforts will strongly enhance thermalization if a more efficient cooling mechanism is
indeed present, since otherwise, in the EP regime, reducing $\dot{Q_0}$ by 5 orders of magnitude will only reduce
$T_{CBT}$ by a factor of ten.

An alternative avenue based on quantum dot CBTs, e.g. in GaAs, might also be rewarding, taking advantage of a much
larger $E_C$ and level spacing $\Delta$. The resulting reduced sensitivity to the environment might allow a single dot
to be used, rather than an array, cooling the reservoirs directly via the WF term, rather than through a long series of
junctions.

\vspace{-4mm}\begin{acknowledgments}\vspace{-4mm} We would like to thank R.~Blaauwgeers, G.~Frossati, R.~Haley,
G.~Pickett, V.~Shvarts, P.~Skyba and A.~de~Waard for very useful discussions. This work was supported by the Swiss
Nanoscience Institute SNI, NCCR QSIT, Swiss NSF, ERC starting grant, and EU-FP7 SOLID and MICROKELVIN.
\end{acknowledgments}

%\bibliography{cbt}

%merlin.mbs 2010-03-15 4.21a (PWD, AO, DPC)

\end{document}